\documentclass[preprint,showpacs,preprintnumbers,amsmath,amssymb]{revtex4}
\usepackage{graphicx}
\usepackage{dcolumn}
\usepackage{bm}

\begin{document}
\draft
\title
{Magneto-Optics in parabolic three-dimensional quantum dots in 
magnetic fields of arbitrary direction}

\author{Mart\'{\i} Pi$^1$, Francesco Ancilotto$^2$, and Enrico Lipparini$^3$}

\address{
$^1$Departament d'Estructura i Constituents de la Mat\`eria,
Facultat de F\'{\i}sica,
Universitat de Barcelona, E-08028, Spain.}
\address{
$^2$INFM (Udr Padova  and DEMOCRITOS National Simulation  Center,
Trieste, Italy)
and Dipartimento di Fisica, Universit\`a di Padova, I-35131 Padova,
Italy}
\address{
$^3$Dipartimento di Fisica, Universit\`a di Trento, and
INFM sezione di Trento, I-38050 Povo, Italy.}

\date{\today}

\begin{abstract}
We generalize the Kohn's theorem to the case of parabolic three-dimensional 
quantum dots in magnetic fields of arbitrary direction.
We show numerically that the exact resonance 
frequencies in the magneto-optical
absorption of these dots are reproduced by the adiabatic time-dependent 
local spin density 
approximation theory (TDLSDA). We use TDLSDA to predict spin density
excitations in the dots.
\end{abstract}

\maketitle
\section{Introduction}

Longitudinal dipole excitations in quantum dots (QD's)
under applied magnetic fields have been observed
in photo-absorption experiments in the far infra-red region,\cite{Sik89,Dem90}
and more recently in Raman scattering
experiments.\cite{Str94,Sch96} These modes are excited by the 
dipole operator ${\vec D}=\sum_j {\vec r}_j$,
and the resonance frequencies in the magneto-optical absorption spectrum 
of a QD with asymmetric
parabolic confinement potential are found to be independent of the 
electron-electron interaction and of the number
of electrons in the dot, and they coincide with
the single-electron transition frequencies. This statement, 
known as the generalized Kohn theorem, was demonstrated some years ago by
Peeters\cite{Pee90} for a 
two-dimensional QD under an external magnetic field ($B$) perpendicular
to the plane of motion
of the electrons. Peeters gave the following explicit formula for the 
resonance frequencies: 

\begin{equation}
\omega_{1,2}^2=\frac{1}{2}\left\{(\omega_x^2+\omega_y^2+\omega_c^2)\pm
\sqrt{(\omega_x^2+\omega_y^2+\omega_c^2)^2-4\omega_x^2\omega_y^2}\right\}~,
\label{eq1}
\end{equation}
where $\omega_c=e B/m c$ is the cyclotron  frequency, and $\omega_x$,
$\omega_y$ are the confinement frequencies in the $x$ and $y$ directions. 
Time-dependent theories as time-dependent Hartree-Fock,
and adiabatic time-dependent local spin density approximation (TDLSDA) 
fulfil the generalized Kohn theorem, \cite{Ser99,Lip99,Ste99}
and thus any numerical implementation of these methods should
reproduce the exact result of Eq. (\ref{eq1}). Viceversa,
static mean field theories like Hartree, Hartree-Fock and Kohn-Sham theories 
violate the theorem.\cite{Lip03}

In this paper we generalize the result given in
Eq. (\ref{eq1}) to the case of a parabolic three-dimensional QD
in a magnetic field of arbitrary direction, and 
present and discuss adiabatic TDLSDA results for the dipole excitations 
of the same system. For density modes, the numerical calculation
reproduces  the exact result, and must be considered
as a stringent test of our three-dimensional (3D) TDLSDA
code in view of its application to study other  excitation modes,
like the spin dipole modes considered here, or to address the far-infrared
response of more complex systems such as vertical diatomic artificial 
molecules.\cite{last}

\section{Generalized Kohn's theorem for 3D quantum dots
in magnetic fields of arbitrary direction}

In the effective mass approximation, the Hamiltonian for $N$
noninteracting electrons confined in a QD 
by harmonic potentials of freqencies $\omega_x, \omega_y$ and
$\omega_z$ in the $x, y$ and $z$ directions, in 
a constant magnetic field ${\vec B}$, is written as

\begin{equation}
H_0=\sum_{j=1}^N h_0(j)=\sum_{j=1}^N\left\{\frac{1}{2m}\left[{\vec p}_j
-{e\over c}{\vec A}_j\right]^2+
\frac{1}{2}m\left(\omega_x^2 x_j^2+\omega_y^2 y_j^2+\omega_z^2 z_j^2\right)
+g^*_s\mu_B{\vec B}\cdot{\vec s}_j\right\}~,
\label{eq2}
\end{equation}
where $m= m^* m_e$ is the electron effective mass,
${\vec A}$ is the vector potential which we write in the symmetric gauge as
${\vec A}=({\vec B}\land{\vec r})/2$, and $g^*_s$ is the effective gyromagnetic 
factor. In the numerical calculations we have used the values of $m^*$,
dielectric constant $\epsilon$
and $g^*$ of GaAs, namely, $m^*= 0.067, \epsilon= 12.4$,
and $g^*= -0.44$.  Without loss of generality, we write 
${\vec B}=B(\sin\theta,0,\cos\theta)$, where $\theta$ is the angle between
the magnetic field and the $z$ axis.  
Using hereafter effective atomic (dot) units (d.u.) 
$\hbar=e^2/\epsilon=m=1$, it can be easily checked that

\begin{equation}
h_0=\frac{1}{2}{\vec p}\,^2+
\frac{1}{2}\left(\omega_x^2 x^2+\omega_y^2 y^2+\omega_z^2 z^2\right)
+h_{m}~,
\label{eq0}
\end{equation}
where
\begin{equation}
\left.
\begin{array}{cr}
h_{m}&=\frac{1}{8}\omega_c^2[x^2 \cos^2\theta+y^2
+z^2\sin^2\theta-2xz\cos\theta\sin\theta]
+\frac{1}{2}g^*_s\mu_B\eta_{\sigma}
\nonumber
\\
&-\frac{i}{2}\omega_c\left[\sin\theta\left(y\frac{\partial}{\partial
z} -z\frac{\partial}{\partial y}\right)
+\cos\theta\left(x\frac{\partial}{\partial y}
-y\frac{\partial}{\partial x}\right)\right]~,
\end{array}
\right.
\label{eq3}
\end{equation}
and $\eta_{\sigma}=+1(-1)$ for $\sigma=\uparrow(\downarrow)$ with respect 
of the direction of $\vec{B}$.
The interacting system is described by the 
Hamiltonian $H=H_0+V$, where
$V$ is the electron-electron interaction
\begin{equation}
V=\sum_{i<j=1}^N\frac{1}{|{\vec r}_i-{\vec r}_j|}~.
\label{eq4}
\end{equation}

The single particle Hamiltonian $h_0$ can be exactly diagonalized:

\begin{equation}
h_0=\sum_{\alpha=1}^3\omega_{\alpha}\left(c^+_{\alpha}c^-
_{\alpha}+\frac{1}{2}\right),
\label{eq5}
\end{equation}
with the creation operator $c^+_{\alpha}$ given by 

\begin{equation}
c^+_{\alpha}=a_{\alpha} x + b_{\alpha} y + c_{\alpha} z + i[d_{\alpha}p_x 
+e_{\alpha}p_y + f_{\alpha}p_z]~.
\label{eq6}
\end{equation}
The operator $c^-_{\alpha}$ is the Hermitian conjugate of
$c^+_{\alpha}$, and the coefficients
$a_{\alpha},...,f_{\alpha}$ in Eq. (\ref{eq6}) are determined by
solving the equation

\begin{equation}
[h_0,c^+_{\alpha}]=\omega_{\alpha}c^+_{\alpha}~,
\label{eq7}
\end{equation}
together with the normalization condition $[c^-_{\alpha},c^+_{\alpha}]=1$.
One gets the following homogenous system of linear equations:

\begin{eqnarray}
&&a \omega
     +\frac{i}{2}  b\omega_c\cos\theta
     + d (\omega_x^2 + \frac{1}{4}\omega_c^2 \cos^2\theta)
     -\frac{1}{4} f \omega_c^2  \sin\theta\cos\theta
    =0 \nonumber\\
&&\frac{i}{2}  a \omega_c\cos\theta
     -b \omega
     -\frac{i}{2}  c \omega_c\sin\theta  
     -e (\omega_y^2 +\frac{1}{4} \omega_c^2)
    =0 \nonumber\\
&&\frac{i}{2}  b \omega_c\sin\theta
     -c \omega 
     + \frac{1}{4} d \omega_c^2\sin\theta\cos\theta 
     -f (\omega_z^2+\frac{1}{4}\omega_c^2\sin^2\theta)
    =0 \nonumber\\
&&a
     +d \omega
     +\frac{i}{2} e \omega_c\cos\theta
    =0 \nonumber\\
&&b
     -\frac{i}{2} d \omega_c\cos\theta
     +e \omega
     + \frac{i}{2} f\omega_c\sin\theta
    =0 \nonumber\\
&&c
     - \frac{i}{2} e \omega_c\sin\theta
     +f \omega
    =0
\label{eq8}
\end{eqnarray}

from which 
the energies $\omega_1, \omega_2, \omega_3$
are obtained
by solving the secular equation ($x\equiv\omega^2$): 

\begin{eqnarray}
&&x^3 -x^2(\omega_c^2+\omega_x^2+\omega_y^2+\omega_z^2)
\nonumber\\
&&-x(\omega_c^2\omega_x^2\cos^2\theta -\omega_c^2\omega_z^2\cos^2\theta
-\omega_y^2\omega_z^2-\omega_x^2\omega_c^2 -\omega_x^2\omega_y^2-\omega_x^2
\omega_z^2)
\nonumber\\
&&-\omega_x^2\omega_y^2\omega_z^2=0~.
\label{eq03}
\end{eqnarray}

For each energy solution $\omega_{\alpha}$, Eqs. (\ref{eq8}) supplemented with
the normalization condition
\begin{equation}
{\rm Re}\left[a d^* + b e^* +c f^*\right]=-\frac{1}{2N}~,
\label{eq01}
\end{equation}
where the $*$ indicates complex conjugation, give
the coefficients $a_{\alpha}$,...,$f_{\alpha}$.

Defining $C^+_{\alpha}= \sum_{j=1}^N c^+_{j,\alpha}$, 
it is easy to prove from Eqs. (\ref{eq4}) and (\ref{eq6}) that 

\begin{equation}
[V,C^{\pm}_{\alpha}]=0~,
\label{eq02}
\end{equation} 
which is valid not only for the 
Coulomb interaction but also for any $V$ that depends only on the relative 
distance between any two electrons.  
From Eqs. (\ref{eq7}) and (\ref{eq02}) it follows immediately that
the Hamiltonian $H=\sum_j h_0(j)+V$ satisfies the equation of motion

\begin{equation}
[H,C^{\pm}_{\alpha}]=\pm\omega_{\alpha}C^{\pm}_{\alpha}~,
\label{eq9}
\end{equation}
which implies that if $|n\rangle$ is an eigenstate of $H$ with energy $E_n$ 
so are  
$C^{\pm}_{\alpha}|n\rangle$ with energies $E_n\pm\omega_{\alpha}$. Since in 
the long-wavelength limit
light photoabsorption is induced  by the dipole transitions, and the dipole
operator can be written
as a sum of $C^+_{\alpha}$ and  $C^-_{\alpha}$ operators, one recovers that
dipole transitions can
only occur from an eigenstate $|n\rangle$ to the eigenstates 
$C^{\pm}_{\alpha}|n\rangle$, and
the  absorption spectrum consists of three peaks of
frequencies $\omega_1, \omega_2$, and $\omega_3$.

As an example, we have solved the homogenous system Eqs. (\ref{eq8})
in the case of an axially symmetric
parabolic QD with $\omega_x=\omega_y=\omega_0$ by taking
$\omega_0=$4.42 meV and $\omega_z$=18 meV,
for some values of the $\theta$ angle. The results for the three
$\omega_{\alpha}$ energies
are reported in Fig. \ref{fig1}. 
Due to our choice of axial symmetry, at $B=0$
the energies $\omega_{1,2}$ are degenerate and equal to $\omega_0$,
and $\omega_3$ coincides with $\omega_z$.
At $\theta=0$, when $\vec{B}$ is parallel to the symmetry axis of the dot, the
solution for $\omega_{1,2}$ coincides with Peeters' expression 
Eq. (\ref{eq1}), and
$\omega_3$ is $B$ independent and equal to $\omega_z$. At
$\theta=90^o$, when $\vec{B}$ is
perpendicular to the symmetry axis of the QD, it is possible to write
down an analytical expression for $\omega_{2,3}$:    

\begin{equation}
\omega_{2,3}^2=\frac{1}{2}\left\{(\omega_0^2+\omega_z^2+\omega_c^2)\pm
\sqrt{(\omega_z^2+\omega_0^2+\omega_c^2)^2-4\omega_z^2\omega_0^2}\right\}~,
\label{eq10}
\end{equation}
whereas $\omega_1$ is $B$ independent and equal to $\omega_0$.

The dipole strength        

\begin{equation}
S_{\bf {\hat e}}(\omega)=\sum_n|\langle
n|\sum_j{\bf {\hat e}}\cdot{\vec r}_j|0\rangle
|^2\delta(\omega-(E_n-E_0))~,
\label{eq11}
\end{equation}
where ${\bf {\hat e}}$ 
is the polarization direction and 
$D_{\bf {\hat e}}=\sum_j{\bf {\hat e}}\cdot{\vec r}_j$  the dipole operator, 
can be also analytically calculated, and is different from zero 
only when the excitation energy
$(E_n-E_0)$ is equal to the frequencies $\omega_{\alpha}$. One can show that
the energy-weighted sum rule $m_1$
\begin{equation}
m_1=\int S(\omega)\, \omega \,d\omega=\sum_n(E_n-E_0)
\left|\left\langle n\left|D_{\bf {\hat e}}\right|0\right\rangle\right|^2
\delta(\omega-(E_n-E_0))={N\over2}~,
\label{eq12}
\end{equation}
is exhausted by the three excited states
at the energies $\omega_{1,2,3}$. 
The $m_{-1}$ sum rule, related with
the static polarizability, can also be worked out yielding a result that
is also ${\vec B}$ independent:
\begin{equation}
m_{-1}=\int S(\omega)\frac{d\omega}{\omega}
=\frac{N}{2}\sum_q\frac{e^2_q}{\omega^2_q},\,\,
q=x,y,z \label{eq12b}
\end{equation}
(see also Ref. \onlinecite{Lip97}).
By changing the values of $\theta$ and
$B$, some strength moves from one state to another, always preserving the
values of $m_{-1}$ and $m_1$.

\section{TDLSDA calculations}

We now turn to our main goal wich is to
numerically study within TDLSDA the dipole
modes both in the density and spin-density channels. 
As we have indicated in the Sec. I, in this approach
the generalized Kohn theorem discussed previously still
holds and the TDLSDA numerical results provide only a test of the 3D+time
TDLSDA code. In the spin density channel 
the  theorem does not hold,
and the calculations provide an alternative prediction 
for this mode to that provided by the explicit
evaluation of the spin-density correlation function.\cite{Ser99}
We recall that spin dipole modes have been experimentally detected in Raman
scattering experiments.\cite{Str94,Sch96}

To obtain the dipole strength in the two channels, we study the time evolution,
following an initial perturbation, of the dipole signal

\begin{equation}
{\cal D}(t)={\bf {\hat e}}\cdot\langle{\vec D}\rangle~,
\label{eq13}
\end{equation}
where ${\vec D}=\sum_j{\vec r}_j$ for density dipole, and
${\vec D}=\sum_j{\vec r}_j\sigma_j^z$  for spin dipole modes,
and $\langle{\vec D}\rangle$ means average over the time dependent
state.
This method has been used in Refs. \onlinecite{Pue99,Ser99b,Val01,Lip02}
to study charge and current modes of two-dimensional QD and quantum 
molecules;  what follows is a generalization to the 3D case.

To calculate ${\cal D}(t)$, we firstly solve the static Kohn-Sham (KS)
equations
\begin{equation}
\begin{array}{l}
\left[-\frac{1}{2} \left( \frac{\partial^2}{\partial x^2}
+ \frac{\partial^2}{\partial y^2}
+ \frac{\partial^2}{\partial z^2} \right)
+ \frac{1}{2}(\omega_x x^2 + \omega_y y^2 + \omega_z z^2)\right.
\nonumber
\\
\left. 
\,\,+  V^H + V^{xc} + W^{xc}\,\eta_{\sigma}
+ h_m \right]
\Psi_{\sigma}(x,y,z) =
\epsilon_{\sigma} \Psi_{\sigma}(x,y,z) \,\, ,
\end{array}
\label{eq14a}
\end{equation}
where the expression in the square brackets is the KS Hamiltonian ${\cal H}_{KS}$.
More precisely, $V^H(x,y,z)$ is the direct Coulomb potential, $V^{xc}={\delta
{\cal E}_{xc}(n,m)/\delta n}\vert_{gs}$, and
$W^{xc}={\delta {\cal E}_{xc}(n,m)/\delta m}\vert_{gs}$
are the variation of the exchange-correlation
energy density ${\cal E}_{xc}(n,m)$ written in terms of the electron
ground state (gs) density $n(x,y,z)$, and of the local spin magnetization
$m(x,y,z)\equiv n^{\uparrow}(x,y,z)-n^{\downarrow}(x,y,z)$.
The exchange-correlation energy has been taken from
Perdew and Zunger,\cite{Per81} and  ${\cal E}_{xc}(n,m)$
has been constructed as indicated in Ref. \onlinecite{Pi01}.
It is worth noticing that if $B\ne0$  the single particle
wave functions $\Psi_{\sigma}(x,y,z)$ are complex and  their real and
imaginary parts are coupled by $h_m$.

The KS equations are solved in a 3D mesh with $\Delta_x = \Delta_y =$ 0.51
d.u., and $\Delta_z =$ 0.135 d.u., by using  11 points
formulas for the differential operators and a fast-Fourier transform 
to solve the Coulomb potential (for more details see Ref. 
\onlinecite{Anci03}).

Appropriate static solutions of the KS equations are then used as initial
conditions for solving the time-dependent KS equations
\begin{equation}
i\frac{\partial}{\partial t}\psi_{\sigma}({\bf r},t)=
{\cal H}_{KS}\,  \psi_{\sigma}({\bf r},t)\, ,
\label{eq14}
\end{equation}

Specifically, to describe the interaction of the system with an external  
dipole field the gs orbitals are slightly perturbed according to 

\begin{equation}
\psi_{\sigma}^{\prime}({\bf r})=U\psi_{\sigma}({\bf r})
\label{eq100}
\end{equation}
with 

\begin{equation}
U=\exp[i\lambda {\bf {\hat e}}\cdot{\vec r}]~,
\label{eq15}
\end{equation}
for the density dipole modes, and

\begin{equation}
U=\exp[i\lambda \eta_{\sigma}{\bf {\hat e}}\cdot{\vec r}]
\label{eq16}
\end{equation}
for the  spin dipole modes.
Eqs. (\ref{eq100}) and (\ref{eq15}) give rise to an initial state in which
all the electrons of the dot have a rigid velocity in an arbitrary direction
${\bf {\hat e}}$. In the spin dipole case, Eqs. (\ref{eq100}) and
(\ref{eq16}) give initially to spin up and spin down 
electrons a rigid velocity field in opposite
directions. The parameter $\lambda$ is taken small enough to keep 
the response of the system in the linear regime.

We have solved  Eqs. (\ref{eq14}) following the method of
Ref. \onlinecite{Kos88}.  One writes

\begin{equation}
\psi_{\sigma}({\bf r},t+\delta t) - \psi_{\sigma}({\bf r},t-\delta t) =
\left[
e^{-i {\cal H}_{KS}\, \delta t} -
e^{ i {\cal H}_{KS}\, \delta t}
\right]
\psi_{\sigma}({\bf r},t)
\label{eq14b}
\end{equation}
and then expands the square bracket in a Taylor series:

\begin{equation}
\psi_{\sigma}({\bf r},t+\delta t) = \psi_{\sigma}({\bf r},t-\delta t) +
2 \sum_{j=0}^{j_{max}}\left(-1\right)^{j+1}\frac{i}{(2j+1)!}
\left(\delta t\, {\cal H}_{KS}\right)^{(2j+1)}
\psi_{\sigma}({\bf r},t)~.
\label{eq14c}
\end{equation}
At $t=0$ we follow the iteration procedure of Ref.
\onlinecite{Flo78} up to seventh order. After the first $\delta t$ 
step, we
use Eq. (\ref{eq14c}), taking in the expansion  $j_{max}=3$.
Typical $\delta t$ are $\sim 2\times10^{-2}$ d.u.
(1 d.u. is $\sim 5.55\times10^{-14}s$).
After 15000 iterations the total energy is conserved with a relative error
smaller than $10^{-11}$. The dipole signal Eq. (\ref{eq13}) 
is calculated every time-step.

In the linear regime the Fourier transform of the dipole signal 

\begin{equation}
{\cal D}(\omega)=\int dt e^{i\omega t}{\cal D}(t)
\label{eq17}
\end{equation}
is directly related to the dipole strength $S_{\bf {\hat e}}(\omega)$ 
Eq. (\ref{eq11}) by

\begin{equation}
S_{\bf {\hat e}}(\omega)=|{\cal D}(\omega)|~.
\label{eq18}
\end{equation}
Hence, a  frequency analysis of ${\cal D}(t)$ provides the absorption
energies and their associate intensities.\cite{Pue99}

A real time simulation of the dipole evolution  is shown in Fig. 
\ref{fig2} for a $N=6$ quantum dot for
$\theta=22.5^o$ and $B=5$ T. The analysis is made for the three components of the
dipole operator in the density channel, and the frequency analysis of the numerical signals
providing the density dipole strength is plotted in Fig. \ref{fig3} for different
values of $\theta$ and $B$. 
In this figure, the energies $\omega_{1,2,3}$ yielded by
the exact calculation of  Sect. II are indicated with arrows. 
The peak energies of the TDLSDA
strength are also reported with circles in Fig. \ref{fig1}.
From these figures one may conclude that our 
3D TDLSDA calculations
reproduces very accurately the exact results of the previous section.
From Fig. \ref{fig3} one may also see how, by changing the values of $\theta$ and
$B$, some strength moves from one mode to another.

Performing the same simulation in the spin channel we obtain the results
shown in Figs. \ref{fig4} and \ref{fig5}. Differently from the
density channel, in the spin channel all the strength is mainly
concentrated in one collective low energy mode.
The peak energy is lower in the spin than in the density channel.
This is due to the character of the
TDLSDA particle-hole residual interaction, which is attractive
but weak in the spin channel, and repulsive and rather strong in the
density channel, shifting the TDLSDA response from the
single-particle one in opposite directions.\cite{Ser99}
Increasing the value of $B$, the spin mode becomes progressively softer.
The TDLSDA predicts
the existence of a spin instability when the energy of the spin mode 
goes to zero at some critical $B$ value\cite{Ser99,Cio02}.
A spin mode with this feature  has been observed in GaAs quantum
dots\cite{Sch96} using Raman spectroscopy, as well as in
quantum wells\cite{Eri99} in a perpendicular 
magnetic field.
In particular, in the quantum well experiment,  evidence for the spin
instability at some values
of electron density and $B$ has been reported.\cite{Eri99} 
The agreement between our calculation and the experimental
findings of Sch\"uller et al. \onlinecite{Sch96} in QD's is
qualitatively good for  low $B$ values 
(see also Ref. \onlinecite{Ser99}). The lack of experimental
observation of  collective spin states 
at larger $B$ values might be considered the signature of the
above mentioned instabilities. However, it cannot be excluded that
Landau damping, which in this energy range is particularly strong,
prevents its experimental observation. Besides,
correlations beyond TDLSDA which are important 
for non-integer filling factors, 
might also quench the spin mode.

\section{Summary}

The dipole and spin dipole responses of a 3D quantum dot
in a magnetic field of arbitray direction have been analyzed 
by means of real time simulations within TDLSDA.
The fragmetation of the strength due to the non circular shape of the QD
and to the $\theta$ angle between the magnetic field $\vec{B}$ and
the axis perpendicular to the plane of motion of the electrons has been studied.
The density dipole strengths splits in three collective states which are exactly found
at the energies predicted by the generalized Kohn theorem. By changing the values 
of $\theta$ and of $B$, some strength moves from one state to another in
a way which could be easily detected. 

The spin dipole strength is mainly exhausted
by a single, soft collective mode whose energy goes to zero as  $B$ increases,
and as the direction of the applied magnetic field goes from perpendicular to 
parallel to the plane containing the dot. For some values of $\vec{B}$,
TDLSDA predicts spin instabilities similar to these observed in Raman
spectroscopy experiments.

A key result from our study is the circumstantial evidence that it is possible to
develop a 3D code implementing TDLSDA 
with a magnetic field in an arbitrary direction
with very high accuracy. This opens
the possibility of addressing the far-infrared response of more complicated
systems, such as double QD's\cite{last} or quantum rings vertically coupled, 
whose description using the density-density or spin-density
correlation functions\cite{Ser99} is prohibitive. Work
along this line is now in progress.

\bigskip

\section{Acknowledgments}
This work has been performed under grants BFM2002-01868  from
MCyT and 2001SGR-00064 from Generalitat of Catalunya, and has been partially
supported by INFM and COFIN 2001. 
Useful and fruitful discussions with Prof. Manuel Barranco are
gratefully acknowledged.

\newpage
\begin{figure}
\caption[]{ Exact dispersion relation for dipole density modes. Filled
circles are from TDLSDA calculations}
\label{fig1}
\end{figure}

\begin{figure}
\caption[]{Real time evolution of the dipole density signal.
Shown are the $x, y$ and $z$-components of the
dipole signal for $\theta=22.5^o$ and $B$=5T.}
\label{fig2}
\end{figure}

\begin{figure}
\caption[]{Strenght function (arbitrary units) for the dipole density response 
correponding to different angles and magnetic fields.}
\label{fig3}
\end{figure}

\begin{figure}
\caption[]{Real time evolution of the dipole spin signal.
Shown are the $x, y$ and $z$-components of the 
dipole spin signal for $\theta=67.5^o$ and $B$=1T.}
\label{fig4}
\end{figure}

\begin{figure}
\caption[]{Strenght function (arbitrary units) for the dipole spin response 
corresponding to different angles and magnetic fields.}
\label{fig5}
\end{figure}

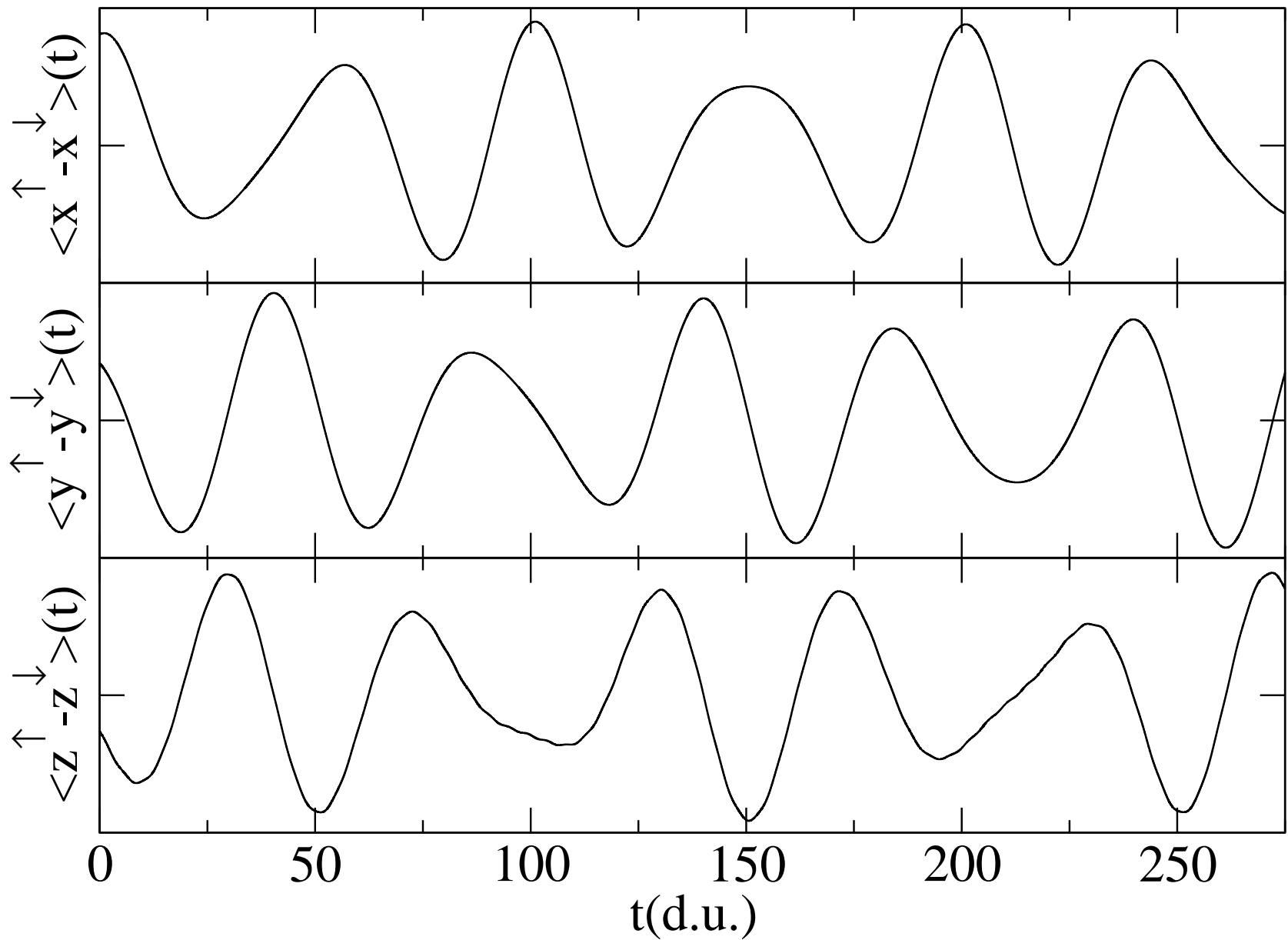
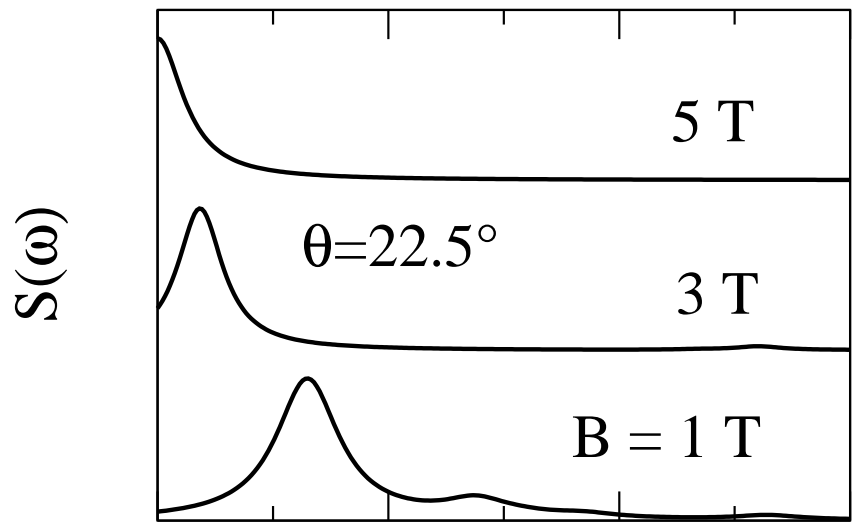
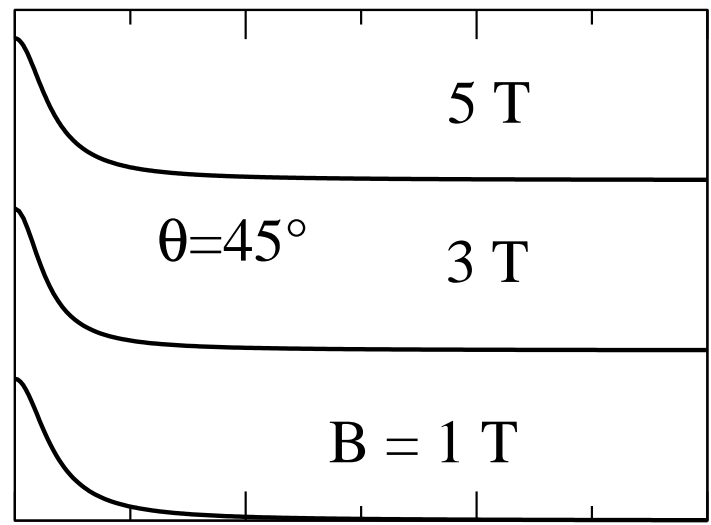
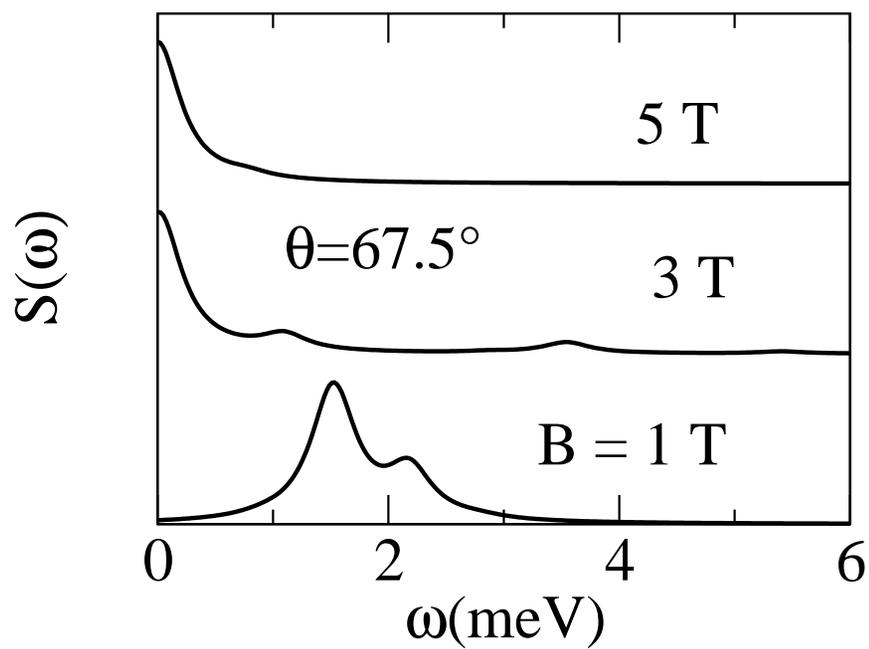
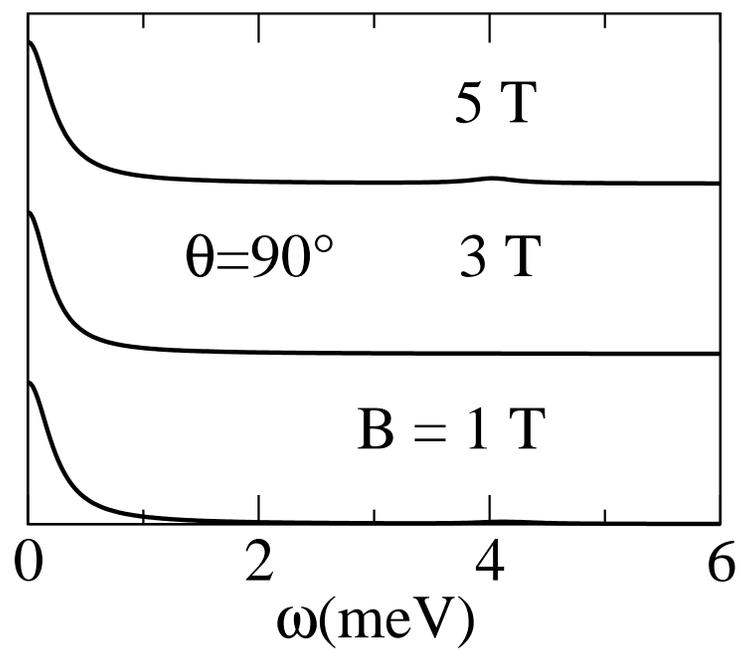
\end{document}